\documentstyle[11pt,paspconf,epsfig]{article}
\def\top#1{\small\it  ADeLA2002 - Astrometry in Latin America\\
    \small\it R. Teixeira, N.V.Leister, P.B-Soares, and V.A.F.Martin  (eds)\\
    \small\it  ADeLA Publications Series, Vol. 1, 2003\small}

\newcommand{\etal}{\mbox{et al.}}

\newcommand{\kms}{\mbox{km s$^{-1}$}}
\newcommand{\muas}{\mbox{$\mu$as}}
\newcommand{\muasyr}{\mbox{$\mu$as~yr$^{-1}$}}
\newcommand{\masyr}{\mbox{mas~yr$^{-1}$}}

\begin{document}
\top{}

\title{VLBI Astrometry}
\author{Norbert Bartel}
\affil{
        York University, Department of Physics and Astronomy, 
        Toronto, ONT M3J 1P3, Canada}

\begin{abstract}
VLBI astrometry allows coordinates of celestial radio sources to be
determined with submilliarcsecond accuracy. In particular, with
differential VLBI astrometry the standard errors of relative positions
and proper motions can be as small as ~$\sim10$ \muas\ and $\sim10$
\muasyr, respectively.  I will review astrometric observations of AGN,
M81$^*$, a supernova, pulsars, and radio stars.
\end{abstract}


\section{Introduction}

The technique of very-long-baseline interferometry (VLBI) provides the
highest accuracy in the determination of coordinates, parallaxes, and
proper motions of celestial sources.  Astrometry surveys of quasars
have established a grid on the sky with position standard errors
smaller than 1 mas. In particular, the International Celestial
Reference Frame (ICRF, Ma \etal\ 1998) with 212 defining extragalactic
reference sources and 396 additional extragalactic sources with a
median position error of only 0.5 mas has been adopted as the fundamental
reference frame of celestial sources. Other surveys as e.g. the VLBA
Calibrator Survey with 1332 extragalactic sources and a median
position error of 0.9 mas (Beasley \etal\ 2002), are tied to the ICRF
and provide important astrometric information for a range of
astronomical and astrophysical projects.  In this paper I will first
describe VLBI observables for astrometry and then summarize results
from differential astrometry, which provides the most accurate
measurements. I will address topics of AGN, supernovae, pulsars, and
radio stars. For astrometric observations of maser sources, see,
e.g., Herrnstein \etal\ (1999) and van Langefeld \etal\ (2000).

\section{VLBI observables for astrometry}
The celestial coordinates of a radio source can be precisely
determined by using the phase observed by an interferometer. Two methods have
been used extensively for phase-sensitive VLBI astrometry, phase-delay
fitting (Shapiro \etal\ 1979; Marcaide and Shapiro ~1983; Bartel
\etal\ 1986) and phase-referenced mapping (e.g., Lestrade \etal\ 1990,
for first uses of astronomical applications, see Gorenstein \etal\
1984 and Bartel \etal\ 1987).  For each interferometer, or baseline, of a
VLBI array the interferometer phase, $\phi(t)$, can be written as
follows:

\vskip -.2in
 
$${{\phi(t)}\over{2\pi\nu}}+{{n(t)}\over{\nu}}=\tau(t)=
\tau_{\rm g}(t)+\tau_{\rm struc}(t)+
\tau_{\rm ion}(t)+\tau_{\rm trop}(t)+\tau_{\rm inst}(t)+\tau_{\rm noise}(t)$$
 
\noindent
where $\nu$~is the radio frequency to which $\phi$ is referred, {\it
n} is the integer number of cycles of phase needed to resolve the
phase ambiguity, and $\tau(t)$ is the phase delay.  On the right side,
$\tau_{\rm g}$ is the contribution to $\tau(t)$ from the geometry of
the interferometer and the fiducial reference point in the source; $\tau_{\rm
struc}$ is that from the source structure about this point; and
$\tau_{\rm ion}$, $\tau_{\rm trop}$, $\tau_{\rm inst}$, and $\tau_{\rm
noise}$ are the contributions, respectively, from the ionosphere, the
troposphere, the instrumentation, and the thermal noise.

Other observables are the fringe rate,
${1\over{2\pi\nu}}{{\delta\phi(t)}\over{\delta t}}$, and the group delay,
${1\over{2\pi}}{{\delta\phi(t)}\over{\delta\nu}}$ .  They can be used directly to
fit the position of the radio source.  The most precise position
determination, however, is obtained by using the phase.  For this
purpose, observations of the target source have to be interleaved with
those of a reference source nearby on the sky. Such differential 
observations will largely remove the effects of the errors of the
terms $\tau_{\rm g}$, $\tau_{\rm ion}$, and $\tau_{\rm trop}$. The
difficulty is to determine the integer, $n$, for each interferometer
phase delay. In phase-delay fitting, this is accomplished iteratively
through ``phase-connection.'' 
When all integers, $n$, are determined, the phase delays are no longer
ambiguous and can be used to estimate, via a weighted least-squares fit,
the position of the radio source. In phase-referenced mapping, the
integers, $n$, are not directly determined but rather  
implicitly through the coordinates of the fiducial reference point 
in the map relative to the coordinates of the reference source.

\section{Close Pairs of AGN}

The first differential astrometry observations were made of the quasar
3C 345 and the (physically) unrelated quasar NRAO 512, separated from
the former by 0.5\arcmin (Shapiro \etal\ 1979).  In later observations,
$\tau_{\rm ion}$ was corrected through simultaneous dual frequency
observations at 2.3 and 8.4 GHz.  The parameter, $\tau_{\rm struc}$,
was corrected through mapping of the sources, and a fiducial point,
the ``core'' of 3C 345, was selected for the position determination.
From essentially two years of observations, the proper motion errors
were as low as $8$ and $21~\muasyr$ in R.A. and dec., respectively.
The core was found to be stationary within 20~$\muasyr$ in R.A., 
equivalent to a bound in velocity at the distance of the source of
0.7$c$. In contrast, the jet components of 3C 345 are moving, largely
along R.A., with an apparent velocity of 14$c$ (Bartel et al.
1986). 

Other observations of AGN pairs with similar separations include the
source Sgr A$^{*}$ in the center of our Galaxy.  Its relative
position and proper motion could be determined with an uncertainty of
0.1~mas and 0.4~\masyr, respectively. This measurement lead to the
conclusion that the orbital velocity of the sun around the
galactic center, assumed to be at a distance of 8~kpc, is 219$\pm$ 20~
\kms (Reid \etal\ 1999).

An important part of VLBI astrometry is the test of predictions of
general relativity (GR). By using two quasars separated by
$\sim$10\deg\ on the sky, Lebach \etal\ (1995) measured the
light deflection from the sun with the highest accuracy obtained so far
and confirmed the GR prediction within 0.1\%. Recently, in similar
measurements of light deflection from Jupiter with three extragalactic sources, 
position uncertainties as
small as $10~\mu$as were obtained (Fomalont \& Kopeikin 2003).

For wider pairs the position error increases since the errors of 
$\tau_{g}$, $\tau_{ion}$, and $\tau_{trop}$
increase.  As an example, the relative positions of the quasar-galaxy
pair 1150+812 - 1803+784 (sep.~15\deg) could be determined with an
error of 0.9 mas (P\'erez-Torres \etal\ 2000). For closer pairs the
position error decreases but is eventually limited by the contribution
from $\tau_{struc}$. For the sources B1342+662 and B1342+663
(sep. 5\arcmin), a relative position with space-VLBI with an
uncertainty of 65~$\mu$as could be obtained (Guirado \etal\ 2001). The 
very close pair of quasars 1038+528 A, B
(sep. 33\arcsec) was monitored for 15 yr and relative core positions with an 
uncertainty of $37~\mu$as were obtained. The upper limit for the proper
motion of one core relative to the other was 10~\muasyr (Rioja \&
Porcas 2000).

\begin{figure}[!t]
\includegraphics{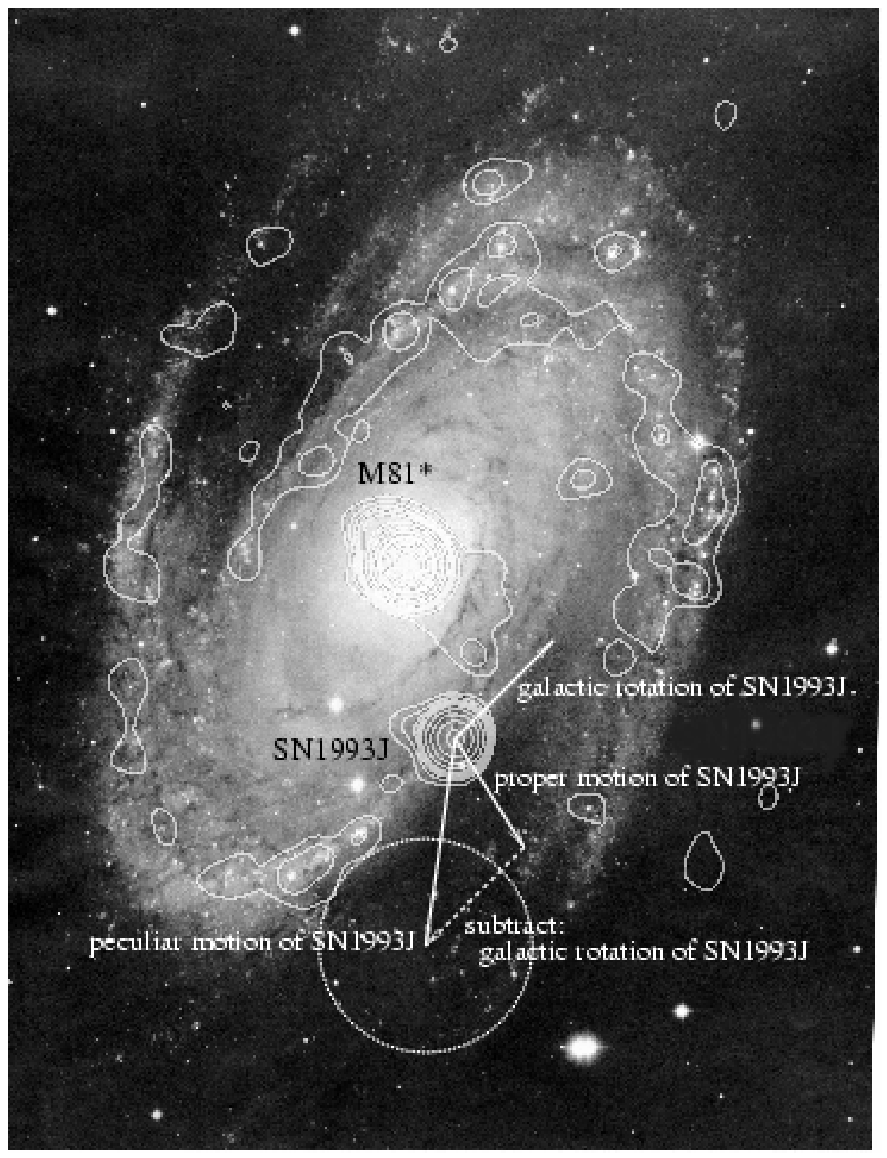}
\includegraphics{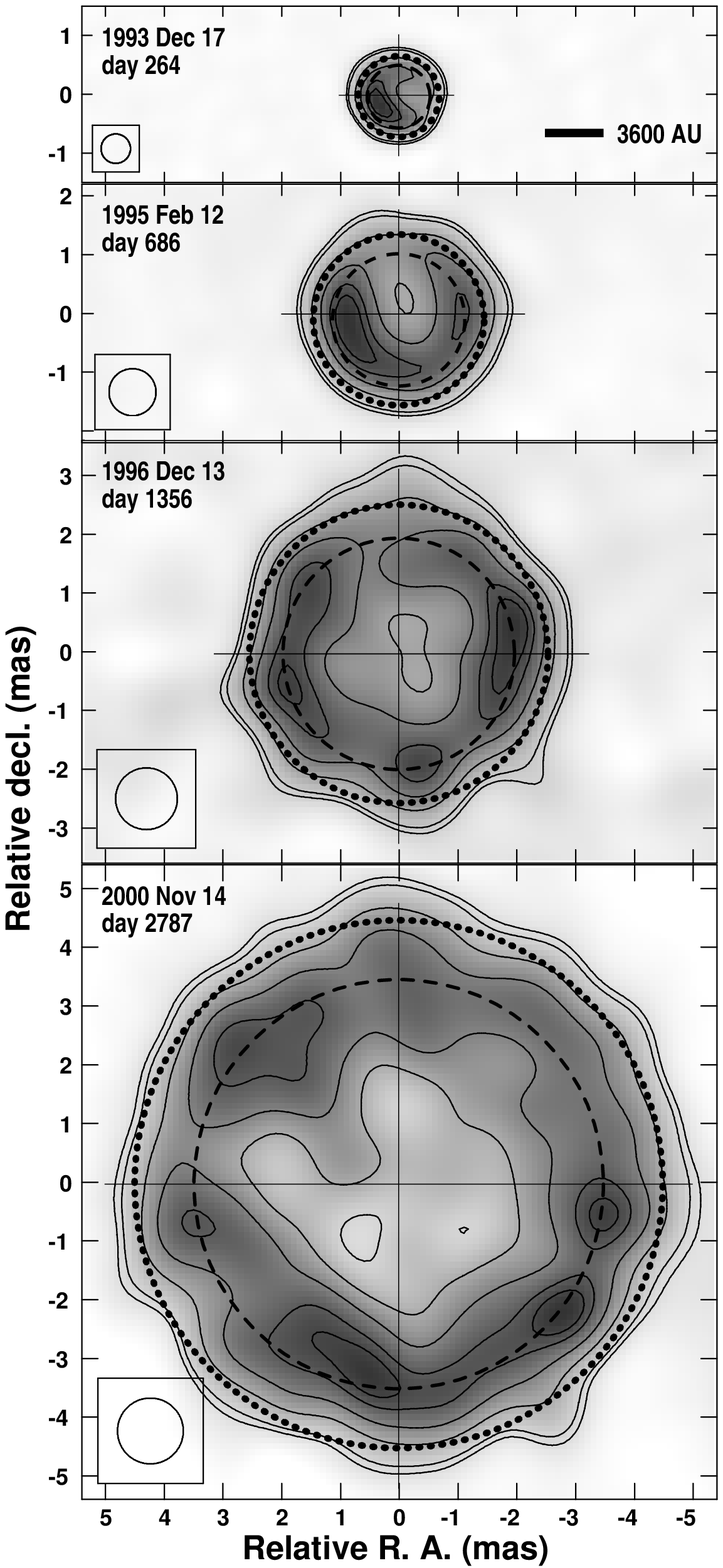}
\vskip 2.6truein
\noindent

\caption[]{Left panel (a): An optical image of M81 (Hubble Atlas),
with overlaid contours showing the radio emission of M81 and SN
1993J. Also shown is the vector of the nominal peculiar motion of the
geometric center of SN 1993J in the co-rotating galactic reference
frame, which is probably not significant (Bietenholz, Bartel, \& Rupen
~2001). Right panel (b): Representative 8.4-GHz VLBI images and models of SN
1993J with the geometric center at the origin (Bartel \etal\ 2002), which is 
within $64~\mu$as (rms) of the explosion center.}
\end{figure}
\begin{figure}[!t]
\plottwo{fig_2a.ps} {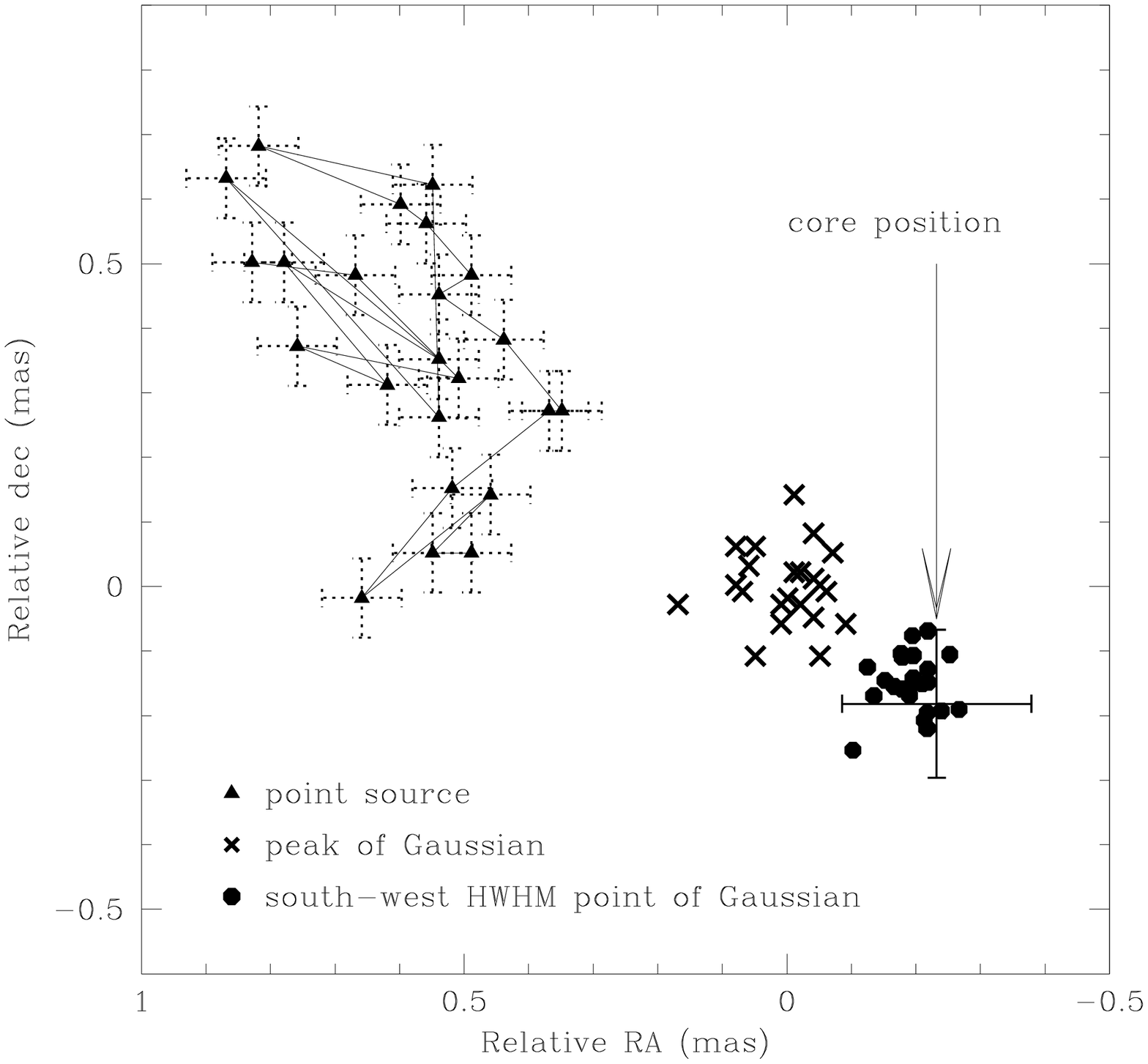} 
\caption[]{Left panel (a): An image of M81$^{*}$ from VLBI
observations at 8.4 GHz. The cross indicates the position of the
stationary core. A jet is emerging toward the northeast.  Right panel
(b): The motion of the jet components (point source, peak of Gaussian)
and the position of the stationary core from 20 epochs relative to the
geometric center of SN 1993J (Bietenholz, Bartel, \& Rupen~ 2000).}
\end{figure}

\section{The Nucleus of M81 and SN 1993J}

SN 1993J in the galaxy M81 is at a distance of 3.63 Mpc one of the
nearest supernovae discovered in the last century and is located
only $\sim$170\arcsec ~away from the galaxy's compact center.  
This center, M81$^{*}$, is, together with that of Cen A, the nearest
extragalactic AGN. With phase-referenced VLBI observations of SN 1993J
and M81$^{*}$ at 24 epochs, from 50 days to several years after the
explosion, the relative coordinates of the explosion center and the
core of M81$^{*}$, presumably the gravitational center of the galaxy,
could be determined with a standard error of 45~\muas, and the nominal
proper motion of the geometric center of the radio shell with a
standard error of 9~\muasyr.  The uncertainties correspond to 160~AU
for the position and 160~\kms\ for the proper motion at the distance
of M81. After correction for the expected galactic proper motion of
the supernova around the center of M81 using HI rotation curves, a
peculiar proper motion of the radio shell center of only $320 \pm
160$~\kms\ to the south was obtained (Fig. 1a), which is probably not
significant. It limits any possible one-sided expansion of the shell
(Fig. 1b) to 5.5\%.

On the other hand, the astrometric measurements give valuable
astrophysical information on the nature of M81$^{*}$. The source is
very compact (Fig. 2a). By identifying the location of its core
(cross) as the most stable point relative to the geometric center of
SN 1993J, a jittery jet directed toward the northeast can be
identified (Fig. 2b), which has a speed up to 0.08$c$. At 8.4 GHz and
particularly at lower frequencies, the core is displaced from the
brightness peak (cf.\ da Silva Neto \etal\ 2002). Because of the
compactness of the source, astrometry proofs to be essential for
studying the kinematics of the jet of this close AGN.

\begin{figure}[!t]
\plottwo{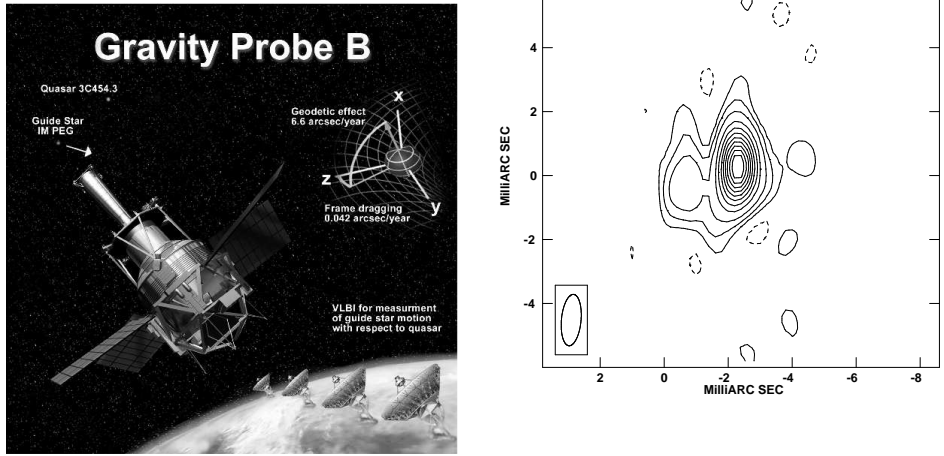} {fig_3b.ps} 
\caption[]{Left panel (a): A poster of the Gravity Probe B mission
with the two predicted GR effects indicated. Right panel (b): An image of
HR 8703 at 8.4 GHz (e.g., Ransom 2003).}
\end{figure}

\begin{figure}[!t]
\plottwo{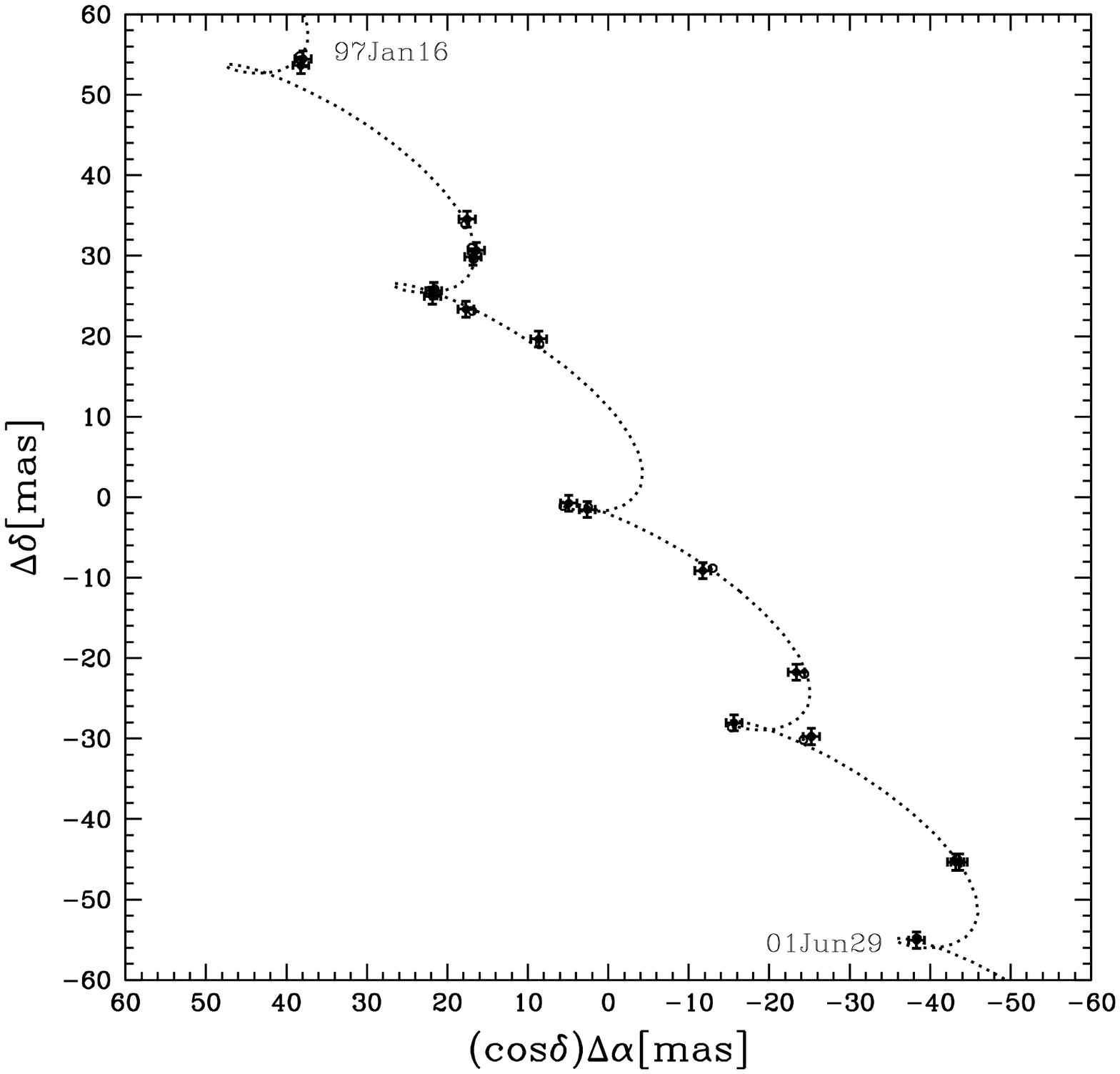} {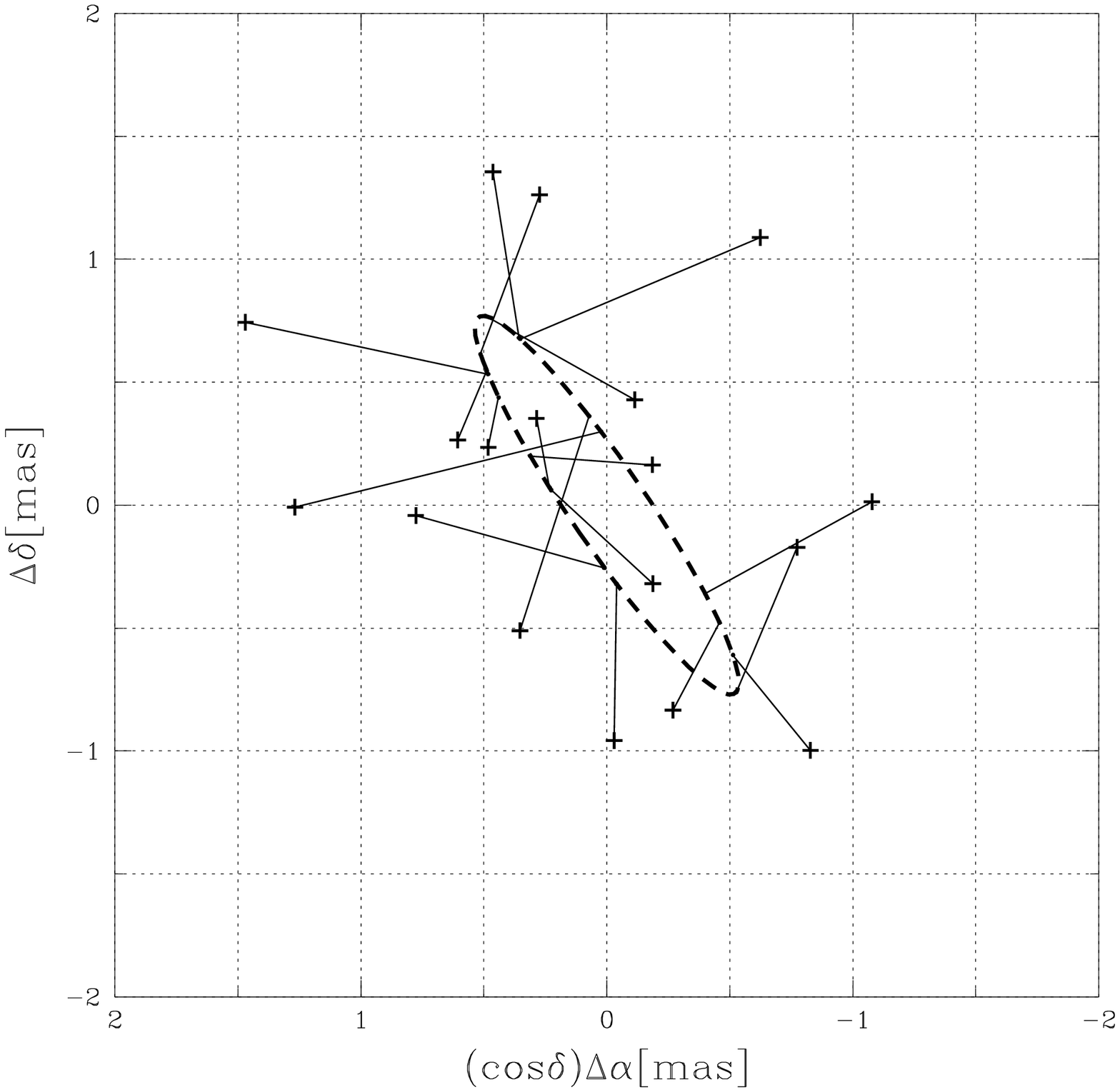} 
\caption[]{Left panel (a): Position determinations of HR 8703 with a
parallax and proper motion fit. Right panel (b): Residual position
measurements with a binary motion fit (e.g., Ransom 2003).}
\end{figure}

\section{Pulsars}
The first astrometric measurement of the position of a pulsar with
VLBI were made by using group delays and phase-delay rates (Bartel
\etal\ 1987) and showed a significant difference to the position
determined with pulse timing (see also Fomalont
\etal\ 1984, for such measurements with the VLA).  It was
caused by the difference between the orientation of the extragalactic
reference frame from that of solar-system dynamic reference frames.
For millisecond pulsars
the precision of pulse timing positions is comparable to, or even
better than, the precision of positions determined with VLBI. A
combination of pulse timing and VLBI observations of millisecond pulsars provided
sufficient information for a tie of the two kinds of reference frames with
milliarcsecond accuracy (Bartel \etal\ 1996; Nunes \& Bartel 1998; cf.\ 
Folkner \etal\ 1994).

A tie between the extragalactic reference frame and the solar-system
dynamic reference frame with an accuracy of $\sim$1~mas could be used
for more accurate interplanetary spacecraft navigation. It could also
be used for a test of GR. Pulse timing of millisecond 
pulsars, combined with the determinations of their positions and proper 
motions with VLBI in the extragalactic reference frame, would make
it possible to measure the annual variation of the gravitational
redshift of earth-bound clocks as predicted by GR with a fractional
uncertainty of $10^{-3}$. Such test would be, by two orders of magnitude,
the most accurate test of the gravitational redshift in the solar potential.

The determination of parallaxes and proper motions provides invaluable
information, e.g., about the birth places of pulsars and the
distribution of free electrons in our Galaxy (e.g., Campbell \etal\
1996; Chatterjee \etal\ 2001; Brisken \etal\ 2002).

\section{Radio Stars}
The first astrometric VLBI measurements of radio stars were made by
using the technique of phase-referenced mapping (Lestrade \etal\
1990). The position determinations were used for a tie with
milliarcsecond accuracy at epoch between the optical Hipparcos
reference frame of stars and the extragalactic reference frame. Other
VLBI measurements resulted in the determination of orbital parameters
(e.g. Guirado \etal\ 1997). The most extensive astrometric campaign
with VLBI of a single star is being conducted for the support of the
NASA/Stanford spaceborne gyroscope relativity mission, Gravity Probe
B. The mission will test the geodetic precession and the frame
dragging effect, predicted by GR, with relative standard errors
smaller than 10$^{-4}$ and 10$^{-2}$, respectively (Fig. 3a). A small
telescope on board the spacecraft is used to lock on to the direction
of the guide star, HR 8703 (IM Peg, Fig. 3b), relative to which the
spin axes motions of four gyroscopes will be determined. The proper
motion of HR 8703 is measured relative to two to three extragalactic reference
sources nearby on the sky with VLBI by using phase-referenced mapping
and phase-delay fitting. These measurements provide the link to
distant inertial space which is essential for the test. Despite HR
8703's structure changes on a scale of 1 mas, the position, parallax,
proper motion, proper acceleration, and binary motion can be
determined, with a standard error, e.g., of the proper motion in 2004
of $\leq 150~\muasyr$ in each coordinate as required for the mission.

\def\nat    {{\sl Nature{\rm}\ }}
\def\prl    {{\sl Phys. Rev. Let.{\rm}\ }}

\vskip 0.15truein
\noindent
For more information on the results from the supernova and Gravity Probe B observations 
and on the collaborators, see: http://aries.phys.yorku.ca/$\sim$bartel/
\end{document}